\documentclass[amsmath,amssymb,aps,twocolumn,pre]{revtex4-2}

\usepackage{graphicx}
\usepackage{dcolumn}
\usepackage{bm}
\usepackage{hyperref}
\UseRawInputEncoding

\usepackage{xcolor}

\begin{document}

\preprint{APS/123-QED}

\title{Flow regimes and repose angle in a rotating drum filled with highly concave particles}

\author{Weiyi Wang \textsuperscript{1}}
\email{weiyi.wang@umontpellier.fr}

\author{Jonathan Barés \textsuperscript{1}}%
\email{jonathan.bares@umontpellier.fr}

\author{Mathieu Renouf \textsuperscript{1}}%
\email{mathieu.renouf@umontpellier.fr}

\author{Emilien Azéma \textsuperscript{1,2,3}}%
\email{emilien.azema@umontpellier.fr}
 
\affiliation{
 \textsuperscript{1} LMGC, Université de Montpellier, CNRS, Montpellier, France
}%
\affiliation{
 \textsuperscript{2} Department of Civil, Geological, and Mining Engineering, Polytechnique Montréal, Montréal, Canada
}%

\affiliation{
 \textsuperscript{3} Institut Universitaire de France (IUF), Paris, France\\
}%

\date{\today}
\begin{abstract}
We present a series of experiments investigating the flow regimes and repose angles of highly concave particle packings in a rotating drum. By varying grain geometry from spherical to highly non-convex shapes, adjusting frictional properties and the particle number of branches, we examine how these parameters and the drum speed influence the flow behavior. Our study identifies two distinct flow regimes: the rolling regime, where granular matter exhibits solid-like behavior near the walls and flows like a liquid near the free surface, and the slumping regime, characterized by cyclic avalanches and solid body rotations. Using quantitative criteria such as the repose angle difference and the area ratio of particle packings, we construct phase diagrams delineating the cross-over between these regimes. Our findings highlight the significant effects of particle concavity, friction, and rotation speed on the flow dynamics of granular materials, providing new insights into the mechanical behaviors of \emph{meta-granular matter}.
\end{abstract}


\maketitle

\tableofcontents

\section{Introduction}
Granular matter has predominantly been studied through the lens of spherical particles. However, grain shape is a crucial factor in understanding the mechanical behavior of aggregates \cite{azema2009_mm,smith2010_PRE,miskin2014_sm,athanassiadis2014_sm,gravish2016_fcsm,zhao2016_gm}. For example, particle angularity has been shown to induce high shear stress and dilatancy \cite{azema2009_mm,azema2007_pre,zhao2019_gm,Binaree_Combineshapefriction_2020}, while particle elongation has been shown to induce nematic ordering, leading to anisotropic flow and bulk properties \cite{azema2010_pre,nagy2023_jsm}.
Some studies have also explored more exotic shapes, such as highly non-convex particles, revealing the unconventional properties of packings made of Z-shaped \cite{murphy2016_gm,karapiperis2022_gm}, U-shaped \cite{franklin2014_epl,marschall2015_gm,karapiperis2022_gm}, and star-shaped or polypod particles \cite{zhao2016_gm,bares2017_epj,zheng2017_epj,zhao2020pre,wang2024_arx,aponte2024_pre}. These studies focus particularly on jamming and stability properties, attempting to explain how particle entanglement can lead to very stable free-standing and even hanging structures.

This new \emph{meta-granular matter} paves the way for more intriguing mechanical behaviors and applications \cite{keller2016_gm,dierichs2016_gm,cantor2022_pp}. Nevertheless, these materials have primarily been studied in their solid phase. Except for a split-bottom Couette shear experiment \cite{mohammadi2022_pre}, little is known about how these very specific granular systems flow.

In the experimental work presented here, sets of particles ranging progressively from spherical to highly concave, and from slippery to highly frictional, flow in one of the most paradigmatic flowing systems, the rotating drum. In this specific loading geometry, provided that the rotation is not too fast and that the entire packing does not slip on the drum edges, it has been shown that the flow can exhibit two distinct regimes \cite{mellmann2001_pt,yang2008_pt,govender2016_me}: (\textit{i}) In the \emph{rolling regime}, the granular matter displays a solid-like behavior near the walls and flows like a liquid near the free surface; (\textit{ii}) In the \emph{slumping regime}, the granular packing cyclically exhibits sequences of avalanches and solid body rotations.

In this paper, we investigate the flow regime of polypod aggregates. Very few studies have focused on the flow of such concave particles, even though they can be present in nature and industry \cite{keller2016_gm,cantor2022_pp}. While our previous work \cite{wang2024_arx} focused on the flow of these particles in the well-behaved rolling regime and for a certain type of particles, here we explore the effects of particle concavity, ranging from very low to very high, their frictional properties, as well as the drum rotation speed.

In the first part, we present our experimental setups and the polypod particles. In the second part, we present results for two specific cases to illustrate the main flow regimes and associated experimental observations, particularly the evolution of the free surface. In the third part, we construct phase diagrams delineating the crossover between slumping and rolling regimes as a function of grain shape, rotation speed, and coefficient of friction. Finally, these results are discussed in the fourth part.

\section{Experimental setup and material}
\begin{figure}[b!]
\centering
\includegraphics[width=0.85\columnwidth]{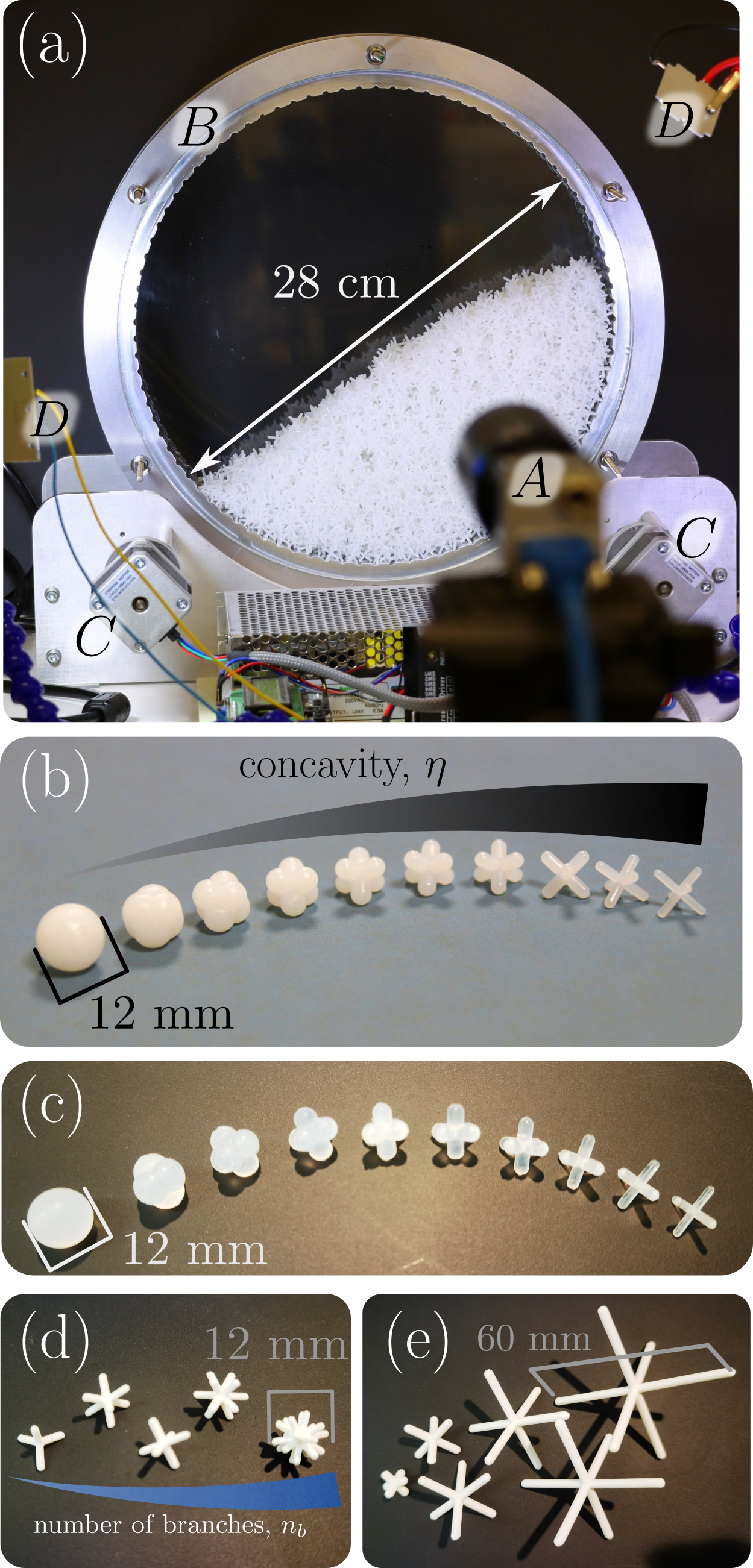}
\caption{(a) Picture of the experimental setup: $A$ represents the camera, $B$ denotes the drum, $C$ indicates the stepper motors, and $D$ represents the lighting system.
(b) HDPE particles and (c) EPDM particles exhibit varying shapes, ranging from spherical ($\eta = 0$) to highly concave ($\eta = 0.87$). As the concavity, $\eta$, increases, the branch diameter gradually decreases from $12$ mm to $1.5$ mm.
(d) PA12 particles exhibit the same concavity ($\eta = 0.87$) with an increasing number of branches, ranging from $n_b=4$ to $n_b=20$.
(e) PP particles showcase a range of shapes, from short branches ($\eta = 0.71$) to long branches ($\eta = 0.94$).}
\label{fig:set_up_grains}
\end{figure}

The experimental setup consists of a rotating drum, as depicted in Fig. \ref{fig:set_up_grains}a, which has been previously described in \cite{wang2024_arx}. This apparatus is a custom-made device comprising a tube with an inner diameter of $28$ cm and variable depth ranging from $5$ cm to $10$ cm. The tube, closed by two glass discs, is mounted on two slick pinions driven by two synchronized stepper motors capable of variable speed control. The system operates within a rotational speed range of $2$ rpm to $5$ rpm, while being recorded at $10$ frames per second (fps) using a $14$ megapixel camera equipped with a $50$ mm lens.

The drum is filled with monodisperse particles of various shapes and materials, as illustrated in Fig. \ref{fig:set_up_grains}b-e. All particles depicted in Fig. \ref{fig:set_up_grains}b-d are inscribed within a sphere of diameter $d=12$ mm. In Fig. \ref{fig:set_up_grains}b-c, they consist of $6$ spherocylinders (also called branches in this paper) extending toward the faces of a regular cube. The radius of the branches, denoted as $r_0$, gradually increases from $0.75$ mm to $6$ mm (sphere case). These particles are fabricated from High-Density PolyEthylene (HDPE) (b) and Ethylene Propylene Diene Monomer (EPDM) (c). The particles depicted in Fig. \ref{fig:set_up_grains}e are of the same nature, composed of PolyPropylene (PP), with a branch section radius of $r_0=3$ mm, while their diameter, $d$, varies from $13$ mm to $63$ mm. All these particles are produced via injection molding using a custom-made mold \cite{wang2024_arx}, generating clusters of particles. In contrast, the particles presented in Fig. \ref{fig:set_up_grains}d are 3D printed using the Slice Laser Sintering (SLS) process with nylon PA12 material. Their diameter and branch section are constant ($d=12$ mm and $r_0=1.5$ mm), but the number of branches, denoted as $n_b$, varies from $3$ to $20$, allowing the branches to emerge from the five platonic solids ($n_b \in [4, 6, 8, 12, 20]$). The choice of materials for the particles ensures they can be considered non-deformable (with Young's moduli around $1$ GPa) and exhibit Coulomb friction coefficients, $\mu$, ranging from $0.2$ (HDPE) to $0.8$ (PA12) (approximately $0.5$ for EPDM, and $0.6$ for PP).

To quantify the particle geometry, a concavity parameter, denoted as $\eta$, is defined as $\eta=(d-2 r_0)/d$. For HDPE and EPDM particles (Fig. \ref{fig:set_up_grains}b-c), this parameter varies such that $\eta \in [0,0.33, 0.5, 0.58, 0.67, 0.71, 0.75, 0.79, 0.83, 0.875]$. For PA12 particles (Fig. \ref{fig:set_up_grains}d), it remains constant at $0.875$, while for PP particles (Fig. \ref{fig:set_up_grains}e), it varies such that $\eta \in [0.77,0.87,0.91,0.93,0.94,0.95]$.

The drum features smooth, transparent glass on both axial sides, while the inner radial side is regularly crenellated to prevent sliding. For HDPE, EPDM, and PA12 particles (see Fig. \ref{fig:set_up_grains}b-d), a constant volume of $900$ mL of particles is loaded into the drum with a depth of $5$ cm. For PP particles (see Fig. \ref{fig:set_up_grains}e), the drum is half-filled with a depth of $10$ cm, since the largest particles are $63$ mm wide. Illumination is provided by LED lights, and the camera is positioned perpendicular to the front glass to capture images of the system. The rotation speed of the drum is controlled by adjusting the speed of two synchronized stepper motors positioned below it (see Fig. \ref{fig:set_up_grains}a). We tested six angular velocities for the drum speed: $\Omega \in [1.93, 2.14, 2.91, 3.20, 4.01, 4.83]$ rpm. For each experiment, the drum is initially rotated for $2$ minutes, followed by a 5-minute recording period. The mean behavior for each set of grain shapes and $\Omega$ is obtained by averaging over $3,000$ frames.

For each frame, the image is thresholded to discriminate the bulk of particles from the rest of the picture. Subsequently, the top of the packing is isolated and described by a polygonal line. This line represents the \emph{free surface}. If this line is regular enough, a slope is fitted at the center of the drum to extract its inclination angle, $\theta$, which corresponds to the \emph{repose angle} when the packing is at rest.

\section{Free surface and repose angle}

When conducting experiments, and as anticipated from the literature \cite{mellmann2001_pt,yang2008_pt,govender2016_me}, two distinct flow regimes are observed depending on the drum speed, $\Omega$, and the exact nature of the particles ($\eta$ and $\mu$). In the simplest case, as previously explored in \cite{wang2024_arx}, the system exhibits a \emph{solid-like} behavior near the walls and deep within the bulk, where particle motion mirrors the solid motion of the drum. Conversely, in the upper part, a liquid-like behavior emerges, characterized by layers of grains exhibiting varying degrees of inertia. This corresponds to the \emph{rolling} regime. In a more intricate scenario, the particle bed undergoes a cyclical process of avalanche and solid body rotation. Specifically, when the bed surface reaches the maximum angle of inclination, a slump occurs during which particles from the upper part of the bed rapidly slide downwards. This corresponds to the \emph{slumping} regime. 

\begin{figure}[b!]
\centering
\includegraphics[width=0.95\columnwidth]{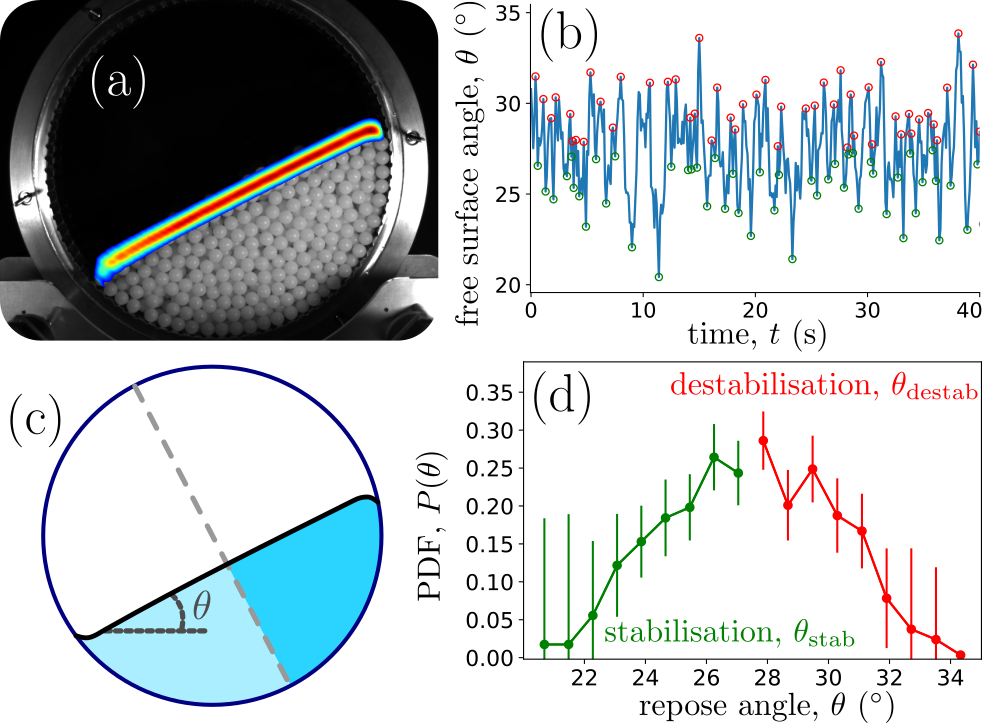}
\caption{Free surface and repose angle for spherical ($\eta=0$) HDPE particles ($\mu \approx 0.2$) with a rotation speed of $\Omega=4.01$ rpm.
(a) Spatial density distribution of the free surface with a frame image as background. Density ranges from blue (low) to red (high).
(b) Evolution of the inclination angle, $\theta$, as a function of time. Red (resp. green) circles indicate destabilization (resp. stabilization) angles.
(c) Cartoon illustrating the packing splitting. The average free surface is represented by a solid black line, with the inclined angle denoted as $\theta$. The gray dashed line perpendicular to the angle, passing through the center of the drum, divides the space occupied by grains into two parts, represented by dark and clear blue zones.
(d) Probability density function of the angle of stabilization, $\theta_{\rm{stab}}$, and destabilization, $\theta_{\rm{destab}}$, represented by the green and red lines, respectively.}
\label{fig:rolling}
\end{figure}

\begin{figure}[b!]
\centering
\includegraphics[width=0.95\columnwidth]{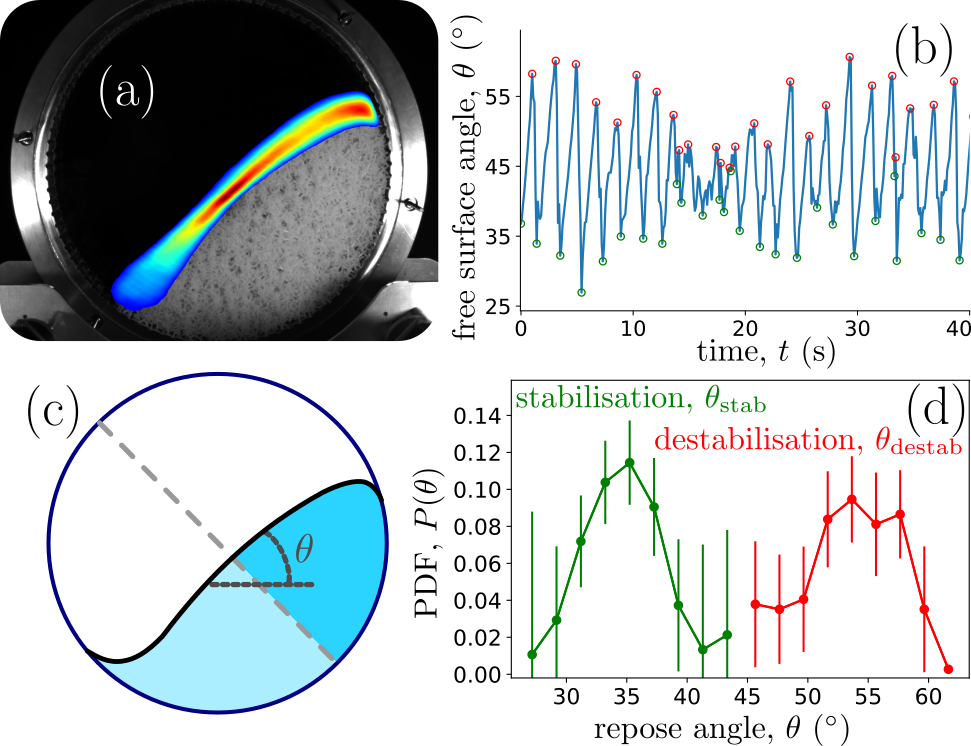}
\caption{Free surface and repose angle for highly concave ($\eta=0.83$) HDPE particles ($\mu \approx 0.2$) with a rotation speed of $\Omega=3.20$ rpm.
(a) Spatial density distribution of the free surface with a frame image as background. Density ranges from blue (low) to red (high).
(b) Evolution of the inclination angle, $\theta$, as a function of time. Red (resp. green) circles indicate destabilization (resp. stabilization) angles.
(c) Cartoon illustrating the packing splitting. The average free surface is represented by a solid black line, with the inclined angle denoted as $\theta$. The gray dashed line perpendicular to the angle, passing through the center of the drum, divides the space occupied by grains into two parts, represented by dark and clear blue zones.
(d) Probability density function of the angle of stabilization, $\theta_{\rm{stab}}$, and destabilization, $\theta_{\rm{destab}}$, represented by the green and red lines, respectively.}
\label{fig:slumping}
\end{figure}

In this paragraph, we present the evolution of the main experimental observables for two significant experiments. Specifically, for HDPE particles ($\mu \approx 0.2$), we demonstrate results obtained with spheres ($\eta = 0$) with a drum rotating at $4.01$ rpm (see Fig. \ref{fig:rolling}), and with highly concave particles ($\eta = 0.83$) with a drum rotating at $3.20$ rpm (see Fig. \ref{fig:slumping}).

For spheres turning rapidly, as depicted in Fig. \ref{fig:rolling}b, concerning the variation of the free surface angle, $\theta(t)$, the rolling regime indicates that $\theta$ gently oscillates around a fixed value with narrow amplitude. Subsequently, by extracting the upper (resp. lower) peaks of this signal, we measure the \emph{destabilization} (resp. \emph{stabilization}) angle of repose, $\theta_{\rm{destab}}$ (resp. $\theta_{\rm{stab}}$). When plotting the Probability Density Functions (PDF) of these quantities, they exhibit a Gaussian-like distribution, as illustrated in Fig. \ref{fig:rolling}d. Additionally, in this regime, the average position of the free surface is characteristic. As demonstrated in Fig. \ref{fig:rolling}a, it forms a homogeneous narrow straight cloud oriented at an angle equal to the average repose angle. Furthermore, in Fig. \ref{fig:rolling}c, a line that cuts the average free surface perpendicularly in its middle divides the particle packing into two distinct regions: dark blue at the top and clear blue at the bottom. Notably, in the specific case of the rolling regime, the areas of both regions are very similar, resulting in the ratio, $\mathcal{R}$, of both regions' area being close to one.

For highly concave particles turning slowly, in Fig. \ref{fig:slumping}, the same quantities as those shown for the rolling regime are plotted for the slumping case. Firstly, as illustrated in Fig. \ref{fig:slumping}b, the free surface angle evolves with a much larger amplitude. The PDFs of the stabilisation-destabilisation repose angles are also distinctive, it is \emph{bimodal}. They form two separate Gaussian-like distributions, with average values clearly distinguished, as depicted in Fig. \ref{fig:slumping}d. Additionally, the average free surface position differs significantly from the rolling regime case. As shown in Fig. \ref{fig:slumping}a, it forms a wider cloud with a wavy backbone. When dividing the packing into two distinct areas as before, it is evident that they are not equal in terms of surface area, resulting in the $\mathcal{R}$ ratio being significantly lower than one.

Key elements in the definition of the flow regime are the free surface and the inclination angle. From the results presented in this paragraph, two quantitative criteria emerge to discriminate between the rolling and slumping flowing regimes. Firstly, we define the repose angle difference, $\Delta \theta$, as the difference between the average stabilisation and destabilisation repose angles: $\Delta \theta = \langle \theta_{\rm{destab}} \rangle - \langle \theta_{\rm{stab}} \rangle$. When the value of this criterion is low, stabilisation and destabilisation angles are close, indicating that the system is more likely in the rolling regime. Conversely, when it is large, stabilisation and destabilisation angles are separated by a significant avalanche phase, suggesting that the system is more likely in the slumping regime.
A second criterion also emerges, the top and lower area packing ratio, $\mathcal{R}$. It is defined as the ratio between the areas separated by a line cutting the average free surface perpendicularly in its middle. If this ratio is close to one, both areas are similar, indicating that the top and bottom packing are balanced, and the system is more likely to be gently rolling. Conversely, if this ratio is significantly lower than one, the average free surface is not straight anymore, suggesting that the system is more likely to be in the slumping regime.

\section{Phase diagram of flow states}
\begin{figure}[b!]
\centering
\includegraphics[width=0.99\columnwidth]{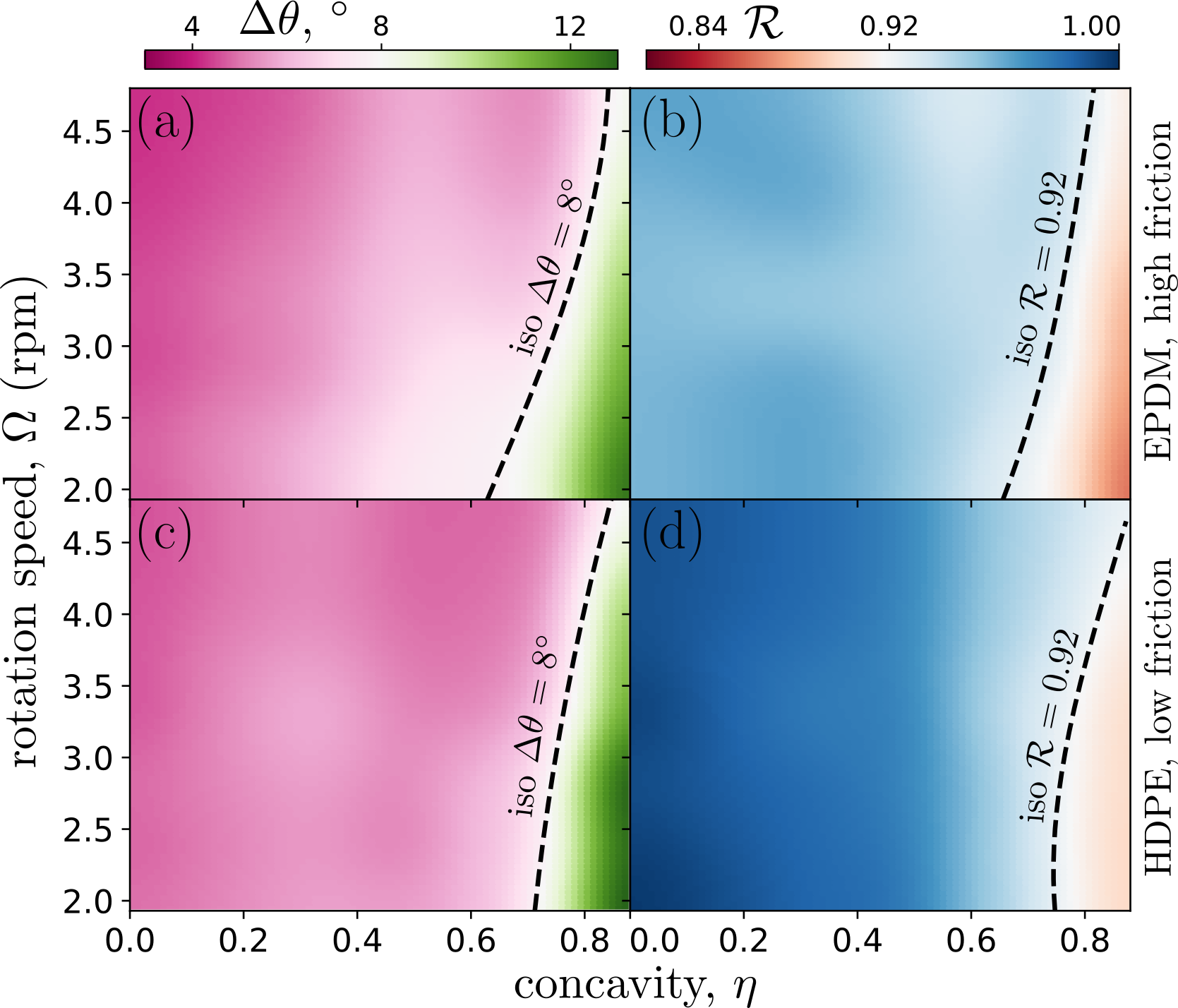}
\caption{Flow phase diagram for high and low friction. Evolution of the flow criteria as a function of the particle concavity, $\eta$, and the drum rotation speed, $\Omega$. The left column represents the $\Delta \theta$ criterion, while the right column represents the $\mathcal{R}$ criterion. The top row is for frictional EPDM particles, while the bottom row is for lower friction HDPE particles. For each phase diagram, a dashed iso-line is plotted at $\Delta \theta = 8$ and $\mathcal{R} = 0.92$, near the regime cross-over.}
\label{fig:diagram_1}
\end{figure}

As discussed in the previous section, two flow regimes are qualitatively observed when particles are rolled into our drum. In parallel, two quantitative observables are constructed, taking very different values for systems in both regimes. In Fig. \ref{fig:diagram_1}a-b, we plot the evolution of these observables, $\Delta \theta$ and $\mathcal{R}$, respectively, for EPDM particles, varying the drum rotation speed, $\Omega$, and the particle concavity, $\eta$. In both figures, an iso-line is plotted for a value in the middle of the full scale. For the first criterion, this is typically $\Delta \theta = 8^{\circ}$, and for the second, $\mathcal{R} = 0.92$.
Firstly, we note that both observables give very similar results, as the phase diagrams resemble each other, and their iso-lines almost superimpose. Additionally, we observe a rapid variation of $\Delta \theta$ or $\mathcal{R}$ around this iso-line, from the minimum value of the observable to its maximum value. This \emph{cross-over} implies that both regimes are significantly distinct: (i) When $\Delta$ is low, around $4$, or $\mathcal{R}$ is high, around $1$, the system is in the rolling regime. (ii) Conversely, when $\Delta$ is high, around $12$, or $\mathcal{R}$ is low, around $0.84$, the system is in the slumping regime. Thus, from Fig. \ref{fig:diagram_1}a-b, we deduce that slumping intermittently occurs when the concavity is high and the rotation speed is low whereas the flow is continuous otherwise in our observation domain.

In Fig. \ref{fig:diagram_1}c-d, we explore the same phase diagram for slippery particles, namely HDPE. The conclusions are similar to those for the top of the figure. However, we observe that when decreasing the interparticle friction coefficient, the slumping regime region shrinks significantly to higher concavity levels and slower rotation speeds. This makes it less likely to observe the intermittent flow regime when particles are less frictional. This tendency is once again observed from both criteria.

\begin{figure}[b!]
\centering
\includegraphics[width=0.70\columnwidth]{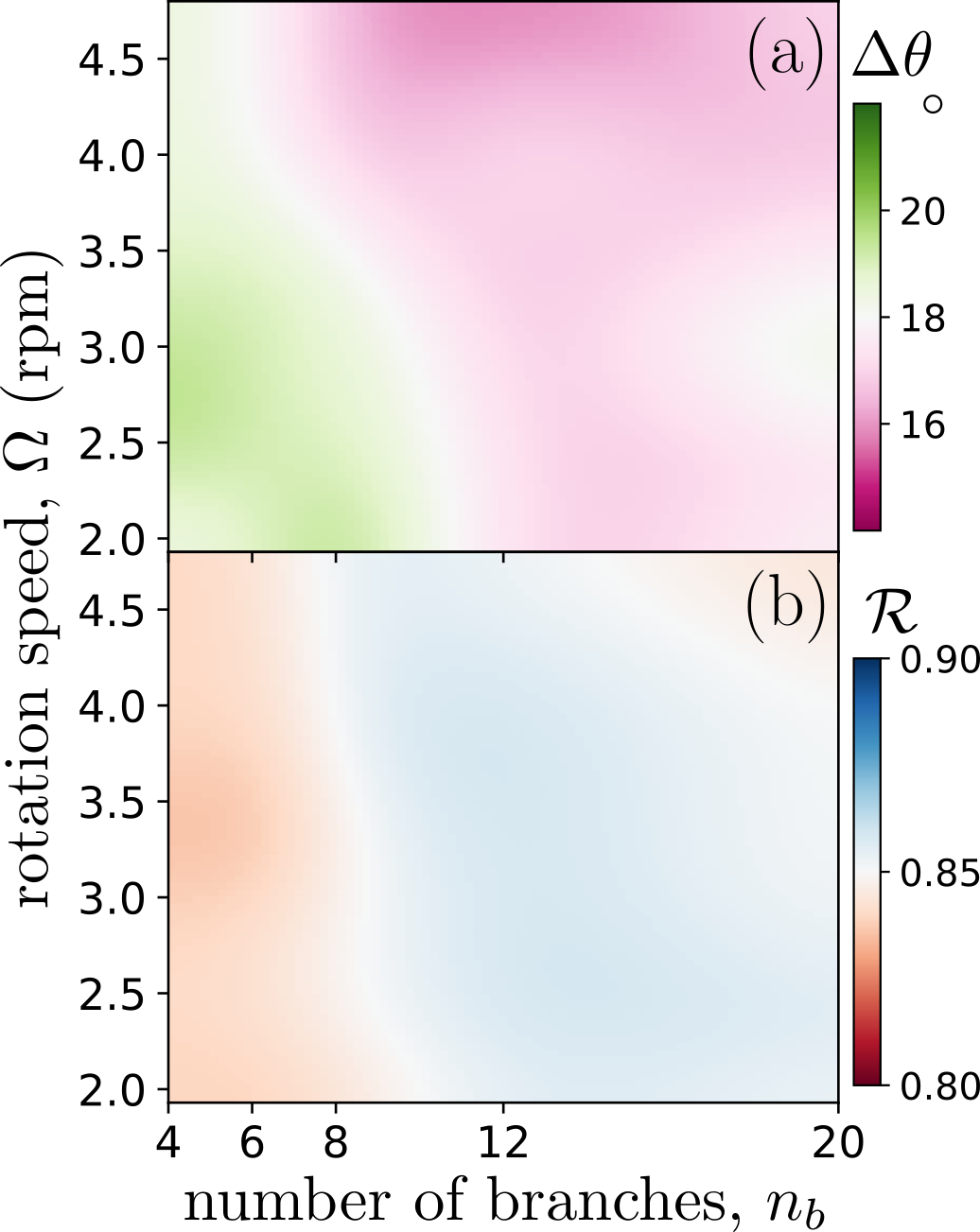}
\caption{Flow phase diagram varying the number of branches for highly frictional particles. Evolution of the flow criteria as a function of the number of particle branches, $n_b$. (a) depicts the $\Delta \theta$ criterion, while (b) illustrates the $\mathcal{R}$ criterion for PA12 particles.}
\label{fig:diagram_2}
\end{figure}

In Fig. \ref{fig:diagram_2}, phase diagrams analogous to those in Fig. \ref{fig:diagram_1} are presented for PA12 particles. As observed previously, increased friction enhances the likelihood of the system entering the slumping regime. Here, PA12 particles are highly frictional, and consistently, the systems are shown to slump intermittently regardless of the number of branches, $n_b$, or the rotation speed, $\Omega$. Indeed, in Fig. \ref{fig:diagram_2}a, $\Delta \theta$ remains above the identified cross-over value of $8^{\circ}$. Similarly, in Fig. \ref{fig:diagram_2}b, $\mathcal{R}$ consistently stays below $0.92$.

Although the system remains within the same regime, discrepancies are observed when varying the rotation speed and the number of particle branches. Depending on these control parameters, regime is more or less prominent. It appears from the phase diagrams in Fig. \ref{fig:diagram_2}a that regime bimodality is more pronounced for packings made of particles with fewer branches and lower rotation speeds. Additionally, in Fig. \ref{fig:diagram_2}b, we observe that the imbalance between the lower and upper parts of the packing is greater when the number of branches is fewer, irrespective of the rotation speed, $\Omega$. However, when $\Omega$ is high, this imbalance is also significant for particles with a high number of branches.

\begin{figure}[b!]
\centering
\includegraphics[width=0.70\columnwidth]{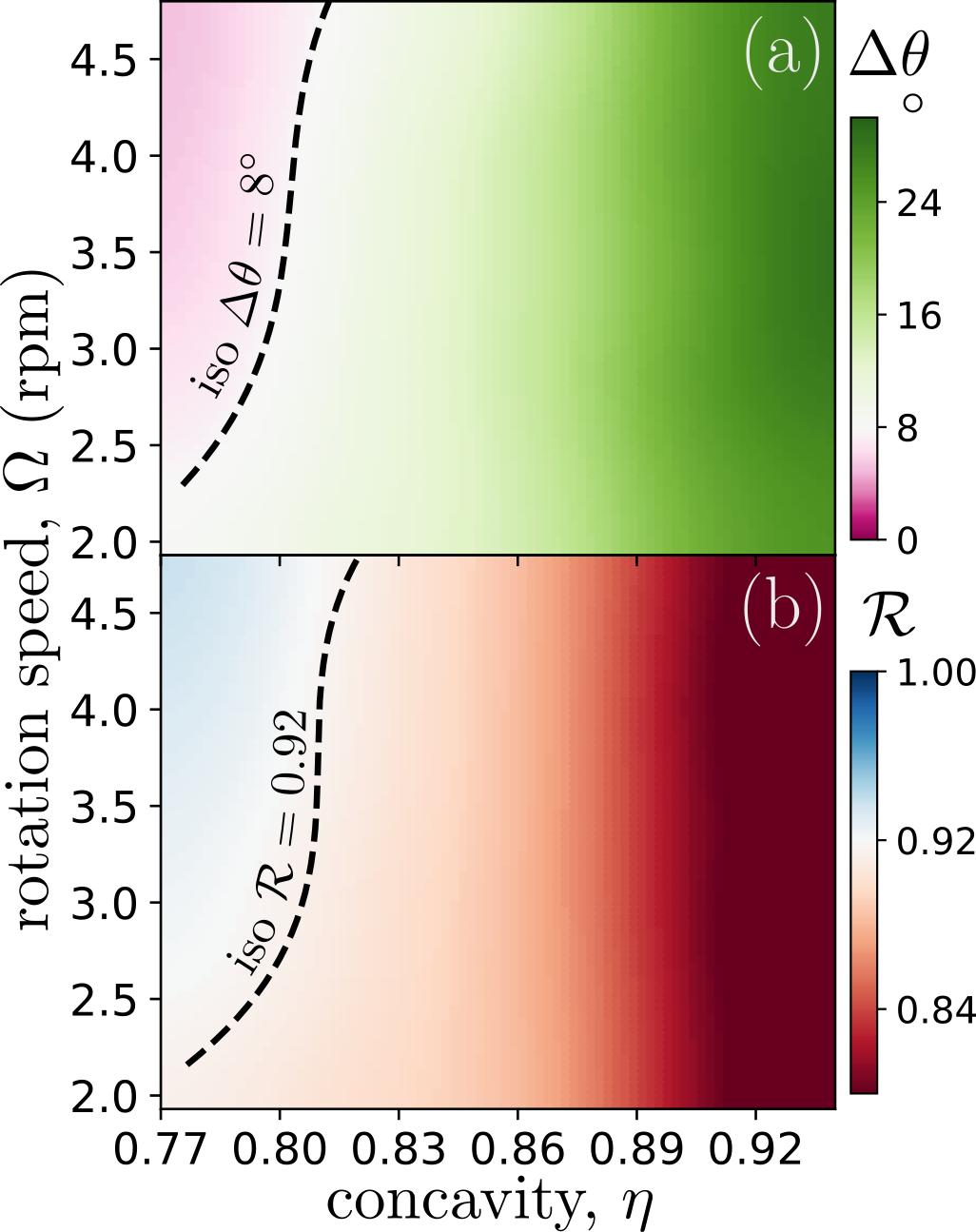}
\caption{Flow phase diagram for highly concave particles. Evolution of the flow criteria as a function of the number of particle particle concavity, $\eta$. (a) depicts the $\Delta \theta$ criterion, while (b) illustrates the $\mathcal{R}$ criterion for PP particles. For each phase diagram, a dashed iso-line is plotted at $\Delta \theta = 8$ and $\mathcal{R} = 0.92$, near the regime cross-over.}
\label{fig:diagram_3}
\end{figure}

In Fig. \ref{fig:diagram_3}, phase diagrams analogous to the previous ones are presented for PP particles. Once again, despite displaying different quantities, these phase diagrams convey the same information. They are also consistent with Fig. \ref{fig:diagram_2}, but allow for a more detailed exploration of the effect of concavity, for higher values. Consistently, we observe that the system is more likely to be in the rolling regime when particle concavity is lower and rotation speed is higher. However, we also note that for high concavity, $\eta$, the effect of the drum rotation speed becomes less significant, as the phase diagrams are vertically invariant on their right-hand side. As $\eta$ increases, we observe that regime bimodality also increases, indicating that higher concavity leads to greater intermittency in the flow regime. No saturation is observed within the range of concavity explored, even though it is quite high.

Based on the foregoing discussion and classification, we have delineated the flow regime of granular materials under various conditions. Notably, for these different experiments, the angle in the rolling regime exhibits relative stability, with fluctuations confined to a narrow range. Thus, it is possible to consider this as a stable angle of repose and to measure its average value, providing insight into the angle evolution for a given experiment (see Fig. \ref{fig:rolling}b). In the slumping regime, the average angle of repose can be measured from the spatial density distribution of the free surface (see Fig. \ref{fig:slumping}a).

\begin{figure}[b!]
\centering
\includegraphics[width=0.85\columnwidth]{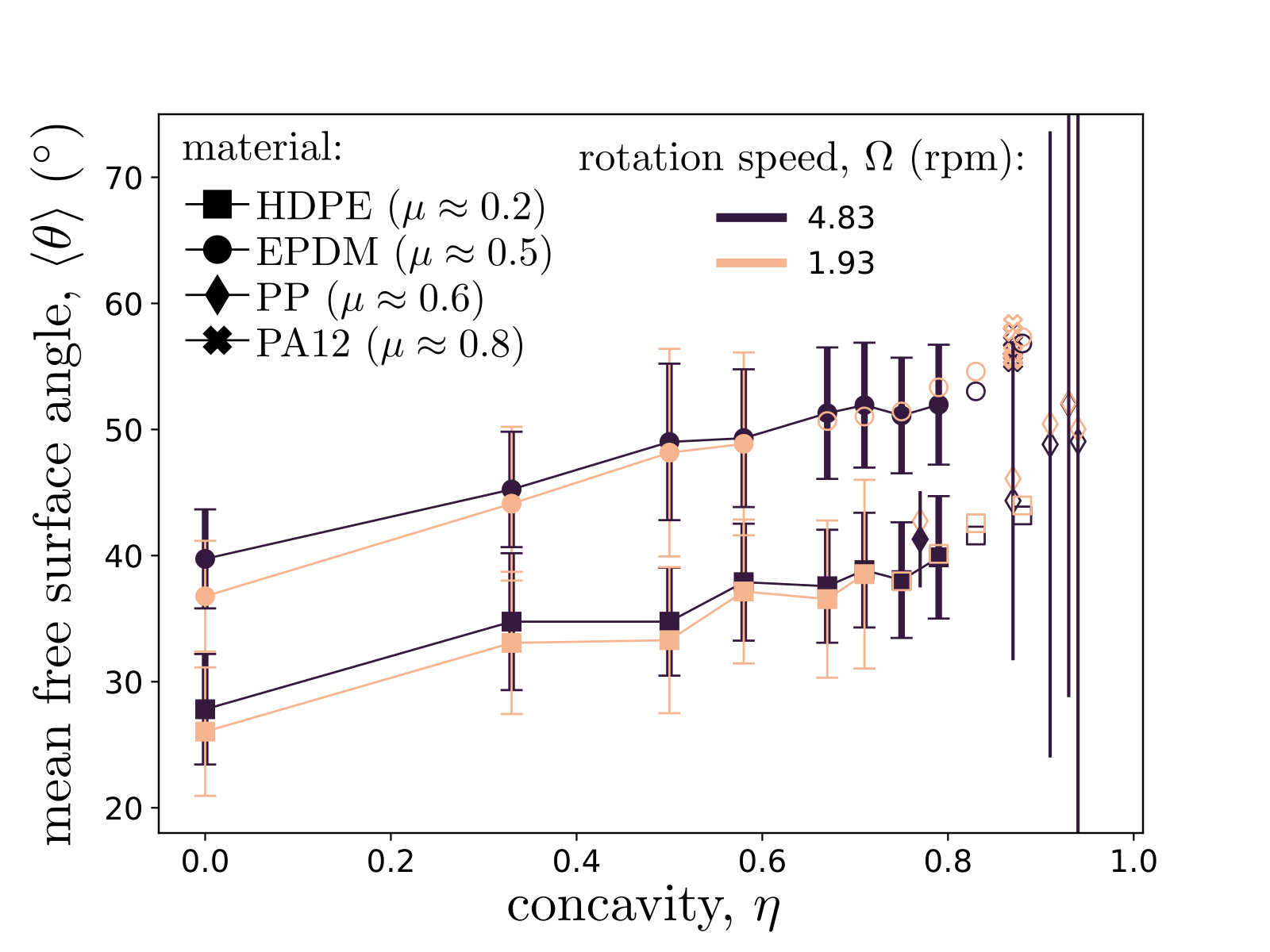}
\caption{Angle of Repose: Variation of the time-average free surface angle $\langle \theta \rangle$ as a function of concavity $\eta$ for different materials of different friction coefficients: HDPE, EPDM, PP, and PA12. Data at maximum drum speed $\Omega=4.83$~rpm are shown in dark, while data at minimum drum speed $\Omega=1.93$~rpm are shown in clear. The solid markers indicate that the flow is in the rolling regime, while the empty markers indicate the slumping regime. Vertical bars show the $95\%$ confidence interval of the average angle.
}
\label{fig:angle_repose}
\end{figure}

In Fig.\ref{fig:angle_repose}, we present variations of the time-averaged angle for all the different particles, for both the maximum and minimum rotation speeds. We observe first that for lower interparticle friction coefficients, the repose angle is significantly lower regardless of the particle shape. Additionally, at low concavity levels, the rotation speed affects the average angle of repose, but this effect diminishes as concavity increases. Specifically, the angle of repose is higher for higher rotation speeds, but only for particle concavity values lower than $\sim 0.7$. It also becomes clear that the angle of repose increases with particle concavity. Below a certain level, this increase appears linear, but for higher concavity values, such as those observed in PP particles, the increase is much steeper.
Even if not presented here, we note that the average angle of repose do not significantly varies with the number of branches. We believe this is mainly due the fact that for PA12 particles behavior is dominated by frictional processes. 

\section{Concluding discussion}
The experiments presented in this study illustrate the rolling and slumping flow regimes of \emph{meta-granular matter} in a rotating drum. The granular systems are composed of star-shaped particles with varying concavity, $\eta$, number of branches, $n_b$, and friction coefficient, $\mu$. This study focuses on the free surface, particularly on the inclination angle when different particles are tested at varying rotation speeds, $\Omega$. Two physical observables are identified to discriminate the flow regime: (\textit{i}) $\Delta \theta$, the difference between the average stabilization/destabilization angles; (\textit{ii}) $\mathcal{R}$, the ratio between the average number of particles at the top and bottom of the drum flow (\textit{cf.} Fig.\ref{fig:rolling}c). In the rolling regime, $\Delta \theta$ is small and $\mathcal{R}$ is close to one, whereas in the slumping regime, $\Delta \theta$ increases while $\mathcal{R}$ decreases. Both observables show similar sensitivity in distinguishing the flow regimes and reveal a crossover between the two regimes when the control parameters ($\eta, n_b, \mu$, and $\Omega$) are adjusted. This provides a quantitative method to evaluate the flow regime by measuring the evolution of the free surface over time. Also, the phase diagrams provide a way to predict granular flow as a function of the grain properties and forcing parameters. 

More precisely, within the observational domain covered by this experimental study, it is found that a flowing system is more likely to be in the rolling regime when (\textit{i}) particle concavity is low, (\textit{ii}) drum rotation speed is high, and (\textit{iii}) interparticle friction is low. Discrepancies have also been observed in the slumping regime when varying the number of particle branches: packing oscillation is higher when the number of branches is either low or high (not intermediate). This observation aligns with previous studies on the stability of such meta-granular matter \cite{aponte2024_pre,aponte2024_arx}, which demonstrated that these particles exhibit \emph{geometric cohesion}. First introduced by Franklin in 2012 \cite{franklin2012_pt}, geometric cohesion describes systems composed of grains with peculiar shapes that interlock and/or mutually entangle, causing the bulk system to behave as a solid even in the absence of adhesive forces between grains. Recent studies \cite{aponte2024_pre} have shown that geometric cohesion for polypods is a non-monotonic function of the number of branches and concavity level. This is due to the fact that a low number of branches does not allow sufficient rotation interlocking, while a high number of branches, given constant concavity, tends to spherical shapes, reducing interlocking capabilities.

Extending the concept of geometric cohesion to systems where external forcing continuously destabilizes the granular system, such as in the drum case, we propose the concept of \emph{geometrical consistency}: Geometrical consistency characterizes systems composed of grains with exotic shapes that entangle, preventing them from easily moving relative to each other. This causes the bulk system to resist flowing, exhibiting properties such as thickness, viscosity, or even stickiness. This study illustrates that for a given interparticle friction coefficient, the ability of particle packing to rearrange in response to external loading and flow varies with particle concavity and the number of branches. It is observed that it is more difficult for a granular system to flow when the concavity is high or when the number of particle branches is either very large or very small. Furthermore, when geometrically induced consistency is high, flow bimodality is also high. This geometrically induced consistency implies a variation in the repose angle, which increases with particle concavity, friction, and the number of branches, provided the number is not too high. This is consistent with observations of geometric cohesion \cite{zhao2016_gm,aponte2024_pre}. Additionally, lower consistency increases the likelihood of the system rolling, whereas higher consistency favors slumping. This is likely because rolling requires particles to rearrange freely and quickly relative to the drum speed, a process hindered by friction or particle interlocking. 

Results obtained with PP particles are consistent with observations for other particles, regardless of changes in the drum width aspect ratio, suggesting limited influence of this geometrical parameter.

The findings reported here represent a significant step forward in understanding the flow of meta-granular systems. They introduce the concept of granular \emph{geometric consistency}, extending \emph{geometric cohesion} to flowing systems. Beyond the scientific implications, this work serves as a “handbook” of mechanical rules for designing granular flow, offering heuristic solutions to customize flow behavior. To the point of view of this handbook, future studies should also explore more complex flowing geometries, other particle shape and also consider the particle softness \cite{bares2022_pp,bares2023_pre} 

\begin{acknowledgments}
The authors acknowledge financial support from ANR MICROGRAM (ANR-20-CE92-0009). We would also like to express our gratitude to Gille Camp, Gilles Genevois, and Quentin Chapelier for their technical assistance in setting up the experimental device. Additionally, we extend our thanks to Arnaud Regazzi, Benjamin Gallard, and Sylvain Buonomo for their assistance in particle preparation. We are grateful to Karen Daniels and Yuchen Zhao for generously sharing the PP particles with us.
\end{acknowledgments}

\bibliography{biblio.bib}

\end{document}